\documentclass[11pt]{article}
\usepackage{amssymb}
\usepackage{amsmath}
\usepackage[all]{xy}
\usepackage{graphicx}

\title{Against Collapses, Purity and Separability\\Within the Definition of Quantum Entanglement\footnote{This paper is an extended version of the discussion and arguments presented in the first part of \cite{deRondeMassri18c}.}}

\author{{\sc C. de Ronde}$^{1,2,3}$ and {\sc C. Massri}$^{4,5}$}
\date{}

\usepackage[margin=2.3cm]{geometry}

\begin{document}

\bibliographystyle{plain}
\maketitle

\begin{center}
\begin{small}
1. Philosophy Institute Dr. A. Korn, University of Buenos Aires - CONICET\\
2. Center Leo Apostel for Interdisciplinary Studies\\Foundations of the Exact Sciences - Vrije Universiteit Brussel\\
3. Institute of Engineering - National University Arturo Jauretche.\\
4. Institute of Mathematical Investigations Luis A. Santal\'o, UBA - CONICET\\
5. University CAECE
\end{small}
\end{center}

\bigskip

\begin{abstract}
\noindent In this paper we  we will argue against the orthodox definition of {\it quantum entanglement} which has been implicitly grounded on several widespread (metaphysical) presuppositions which have no relation whatsoever to the formalism of QM. We will show how these presuppositions have been introduced through a naive interpretation of the quantum mathematical structure which assumes dogmatically that the theory talks about ``small particles'' represented by {\it pure states} (in general, superpositions) which suddenly ``collapse'' when a measurement takes place. In the second part of this paper we will present a non-collapse approach to QM which makes no use whatsoever of particle metaphysics, escaping the need to make reference to {\it space-time separability} or the restriction to {\it certain predictions} of definite valued binary properties. Our paper ends up concluding the essential need to redefine the notion of quantum entanglement, at least in the cases of: i) ``non-collapse'' interpretations of QM; or, ii) any other interpretation which abandons the idea that QM makes reference to ``small particles''.
\end{abstract}
\begin{small} 

\end{small}

\newtheorem{theo}{Theorem}[section]
\newtheorem{definition}[theo]{Definition}
\newtheorem{lem}[theo]{Lemma}
\newtheorem{met}[theo]{Method}
\newtheorem{prop}[theo]{Proposition}
\newtheorem{coro}[theo]{Corollary}
\newtheorem{exam}[theo]{Example}
\newtheorem{rema}[theo]{Remark}{\hspace*{4mm}}
\newtheorem{example}[theo]{Example}
\newcommand{\proof}{\noindent {\em Proof:\/}{\hspace*{4mm}}}
\newcommand{\qed}{\hfill$\Box$}
\newcommand{\ninv}{\mathord{\sim}} 
\newtheorem{postulate}[theo]{Postulate}

\bigskip

\section*{Introduction}

The notion of {\it entanglement} plays today the most central role in what might be regarded to be the origin of a new groundbreaking technological era. In the specialized literature, this research field falls under the big umbrella of what is called {\it quantum information processing}. This term covers different outstanding non-classical technical developments such as, for example, quantum teleportation, quantum computation and quantum cryptography. Without exception, all these new technologies are founded on the notions of {\it quantum superposition} and {\it entanglement}. Even though today, technicians, computer scientists, cryptographers and engineers are rapidly advancing in the creation of new technological devices and algorithms, it is interesting to notice that the notion of entaglement remained almost unnoticed for half a century since its coming into life in 1935 when Albert Einstein, Boris Podolsky and Nathan Rosen discussed what would later become known as the famous EPR {\it Gedankenexperiment}. In that same year, Erwin Schr\"odinger in a series of papers gave its name to the new born concept. He discussed in depth what he called {\it entanglement} ({\it Verschr\"rankung} in German) and showed through the now famous `cat paradox' how quantum correlations rapidly expanded into the classical domain. This new concept was regarded with horror by the community of physicists. ``Spooky'' was the name they invented to bully the still very young notion of entanglement. But after half a century, with the new technical possibilities, what Einstein, Podolsky, Rosen and Schr\"odinger had critically imagined, was now verifiable in the lab. The results of experiments confirmed, against their classical prejudices, the predictions of the theory of quanta. Entanglement grow then rapidly, becoming one of the key concepts used within today's quantum technology. And as in the story of Hans Christian Andersen, the ugly duckling became the most powerful swan. 

However, still today, the concept of entanglement has deep difficulties regarding its definition. In this paper we will argue that the reason behind these difficulties are related to a more fundamental problem within the definition of entanglement itself. In order to understand how this happened we need some history. Since its origin in 1935, the concept was related to two main famous metaphysical presuppositions, introduced within the axiomatic formulation of QM in a completely {\it ad hoc} manner. The first idea is there exists a ``collapse'' of the quantum wave function each time we measure a quantum system (the projection postulate). The second idea ---also completely foreign to the mathematical formalism of the theory--- was that, since QM makes reference to ``small particles'', the principle of separability can be used in order to analyze quantum phenomena. In this paper we attempt to criticize these metaphysical presuppositions on which the notion of entanglement has been built. Our conclusion will be that any interpretation of QM which abandons the existence of collapses and the reference to ``small particles'' will necessarily impose the need to create a new definition of quantum entanglement. 

The paper is organized as follows. In section 1, we go back to the original definitions of separability and entanglement presented in 1935 by Einstein, Podolsky, Rosen \cite{EPR}. In section 2, we present the contemporary definition of quantum entanglement and show how the metaphysics of particles has played a fundamental role by allowing to introduce the notion of pure state as well as the principle of separability. Section 3 discusses the possibilities to advance, beyond the classical metaphysics of particles, taking as a fundament of reasoning only the orthodox quantum formalism. In section 4, we conclude the necessity of producing a new definition of quantum entanglement, at least in the cases of: i) non-collapse interpretations of QM; and ii) interpretations which do not make use of the metaphysics of particles.  Finally, in section 5, we present the conclusions of the paper.

\section{The Origin of Entanglement: EPR's {\it Gedankenexperiment}}

Critical thought is, above all, the possibility of analysis of the foundation of thought itself. The analysis of the conditions under which thinking becomes possible. By digging deeply into the basic components of thinking, one is able to understand the preconditions and presuppositions which support the architecture of argumentation itself. At the heart of the orthodox definition of quantum entanglement lies the analysis of the EPR {\it Gedankenexperiment} presented by Einstein, Podolsky and Rosen in \cite{EPR}.  A critical analysis of the mentioned argument has been of course already provided within the foundational literature ---between many others--- by Diederik Aerts \cite{Aerts84a, Aerts84b} and Don Howard \cite{Howard85}. In the following, we attempt to extend this analysis paying special attention to the notion of separability and to the famous definition of {\it element of physical reality} in order to approach, later on, the presuppositions involved within definition of quantum entanglement.   

\subsection{Einstein's (Spatio-Temporal) Separability Principle}

Even though Albert Einstein was certainly a revolutionary in many aspects of his research, he was also a classicist when considering the preconditions of physical theories themselves. His dream to create a unified field theory was grounded in his belief that physical theories, above all, must always discuss in terms of specific situations happening within space and time. In this respect, the influence of transcendental philosophy in Einstein's thought cannot be underestimated \cite{Howard94}. That space and time are the {\it forms of intuition} that allow us to discuss about objects of experience was one of the most basic {\it a priori} dictums of Kantian metaphysics, difficult to escape even for one of the main creators of relativity theory. In a letter to Max Born dated 5 April, 1948, Einstein writes:
\begin{quotation}
\noindent {\small ``If one asks what, irrespective of quantum mechanics, is characteristic of the world of ideas of physics, one is first stuck by the following: the concepts of physics relate to a real outside world, that is, ideas are established relating to things such as bodies, fields, etc., which claim a `real existence' that is independent of the perceiving subject ---ideas which, on the other hand, have been brought into as secure a relationship as possible with the sense-data. It is further characteristic of these physical objects that they are thought of as arranged in a space-time continuum. An essential aspect of this arrangement of things in physics is that they lay claim, at a certain time, to an existence independent of one another, provided these objects `are situated in different parts of space'. Unless one makes this kind of assumption about the independence of the existence (the `being-thus') of objects which are far apart from one another in space ---which stems in the first place in everyday thinking--- physical thinking in the familiar sense would not be possible. It is also hard to see any way of formulating and testing the laws of physics unless one makes a clear distinction of this kind.'' \cite[p. 170]{Born71}}
\end{quotation}

\noindent This precondition regarding objects situated in different parts of space can be expressed, following Howard \cite[p. 226]{Howard89}, as a principle of spatio-temporal separability:

\smallskip
\smallskip

\noindent {\it {\bf Separability Principle:} The contents of any two regions of space separated by a non-vanishing spatio-temporal interval constitute separable physical systems, in the sense that (1) each possesses its own, distinct physical state, and (2) the joint state of the two systems is wholly determined by these separated states.}

\smallskip
\smallskip

\noindent In other words, the presence of a non-vanishing spatio-temporal interval is a sufficient condition for the individuation of physical systems and their associated states. Everything must ``live'' within space-time; and consequently, the characterization of every system should be discussed in terms of {\it yes-no questions} about physical properties. But, contrary to many, Einstein knew very well the difference between a conceptual presupposition of thought and the conditions implied by mathematical formalisms. In this respect, he also understood that his principle of separability was {\it only for him} a necessary metaphysical condition for doing physics. More importantly, he was aware of the fact there was no logical inconsistency in dropping the separability principle in the context of QM. At the end of the same letter to Born he points out the following:
\begin{quotation}
\noindent {\small ``There seems to me no doubt that those physicists who regard the descriptive methods of quantum mechanics as definite in principle would react to this line of thought in the following way: they would drop the requirement for the independent existence of the physical reality present in different parts of space; they would be justified in pointing out that the quantum theory nowhere makes explicit use of this requirement.'' \cite[p. 172]{Born71}}
\end{quotation}

\noindent This passage shows that Einstein was completely aware of the fact that QM is not necessarily committed to the metaphysical presupposition of space-time separability. But let us now turn to the kernel of the EPR argument, namely, their discussion regarding the sufficient conditions for defining what must be considered as physically real.

\subsection{EPR's {\it Sufficient Condition} for Physical Reality}

The notion of physical reality is of course the key element within the EPR line of argumentation and reasoning. Already in the first page of the paper, \cite{EPR}, Einstein, Podolsky and Rosen introduce the following ``reality criterion'' which stipulates a sufficient condition for considering an element of physical reality:\footnote{We are thankful to Prof. Don Howard for pointing us the specificity of the reality criterion.}

\smallskip
\smallskip

\noindent {\it {\bf Element of Physical Reality:} If, without in any way disturbing a system, we can predict with certainty (i.e., with probability equal to unity) the value of a physical quantity, then there exists an element of reality corresponding to that quantity.}

\smallskip
\smallskip

The relation drawn by the criterion is that between a {\it certain prediction} on the one hand, and the value of a physical quantity (or property) of the system on the other. Certainty is understood as ``probability equal to unity''. Notice that this remark is crucial in order to filter the predictions provided by QM. Only those related to probability equal to one, $p = 1$, can be considered to be related to physical reality. This means, implicitly, that the rest of the quantum mechanical probabilistic predictions which are not equal to one ---namely, those which pertain the interval between 0 and 1---, $p \in (0,1)$, are simply not considered. Given a quantum state, $\Psi$, there is only one meaningful operational statement (or property) that can be predicted with certainty. This leads to the conclusion that only one property can be regarded as being {\it actual}. The rest of quantum properties are considered as being {\it indeterminate}. The important point is that the ``non-certain'' predictions are not directly related to physical reality. Unlike real actual properties, indeterminate properties are considered as being only ``possible'' or ``potential'' properties; i.e., properties that might become actual in a future instant of time (see for a detailed analysis \cite{Sudbery16}). Until these properties are not actualized they remain in an unclearly defined limbo, in the words of Heisenberg \cite[p. 42]{Heis58}, they stand ``in the middle between the idea of an event and the actual event, a strange kind of physical reality just in the middle between possibility and reality.'' The filtering of indeterminate properties ---something which, from an operational perspective, seems completely unjustified---, is directly related to the actualist spacio-temporal (metaphysical) understanding of physical reality which Einstein so willingly wanted to retain. As he made the point \cite{Howard17}:  ``that which we conceive as existing (`actual') should somehow be localized in time and space. That is, the real in one part of space, $A$, should (in theory) somehow `exist' independently of that which is thought of as real in another part of space, $B$.'' As we discussed in detail in \cite{deRondeMassri18a}, it is not so difficult to see ---if we dig a bit deeper--- that this actualist understanding of existence is grounded in the classical representation of physics provided in terms of an {\it actual state of affairs} and {\it binary valuations}. 

But, as noticed by Bohr himself in his famous reply to EPR \cite{Bohr35}, it is the first part of the definition which introduces a serious ``ambiguity''. Indeed, the previous specification, {\it ``If, without in any way disturbing a system,''} refers explicitly to the possibility of measuring the system in question. It thus involves an improper scrambling between ontology and epistemology, between physical reality and measurement. A scrambling ---let us stress---, completely foreign to all classical physics. This scrambling, might be regarded as one between the many ``quantum omelettes'' created during the early debates of the founding fathers \cite[p. 381]{Jaynes}. In fact, this criteria contradicts one of the most interesting characterizations of physical theories provided by Einstein himself. Indeed, according to Einstein \cite[p. 175]{Dieks88a} : ``[...] it is the purpose of theoretical physics to achieve understanding of physical reality which exists independently of the observer, and for which the distinction between `direct observable' and `not directly observable' has no ontological significance''. This is of course, even though ``the only decisive factor for the question whether or not to accept a particular physical theory is its empirical success.'' The physical representation of a physical theory is always {\it prior} to the possibility of epistemic inquiry of which `measurement' is obviously one of its main ingredients. Recalling Einstein's famous remark to Heisenberg \cite[p. 63]{Heis71}: ``It is only the theory which decides what can be observed.''

\subsection{Collapses and Spooky Actions at a Distance}

Today, classical texts that describe QM axiomatically begin stating that the mathematical interpretation of a quantum system is a Hilbert space, that pure states are represented by rays in this space, physical magnitudes by self-adjoint operators on the state space and that the evolution of the system is ruled by the Schr\"{o}dinger equation. Possible results of a given magnitude are the eigenvalues of the corresponding operator obtained with probabilities given by the Born rule. In general the state is mathematically represented as a linear superposition of eigenstates corresponding to different eigenvalues of the measured observables.  Since it is argued that  ``we never observe superpositions'', the theory requires an extra postulate which provides the missing link between quantum superpositions and measurement outcomes. This requirement was explicitly considered by John von Neumann and Paul Dirac in their famous books at the beginning of the thirties. As von Neunmann's \cite[p. 214]{VN} made the point: ``Therefore, if the system is initially found in a state in which the values of $\mathcal{R}$ cannot be predicted with certainty, then this state is transformed by a measurement $M$ of $\mathcal{R}$ into another state: namely, into one in which the value of $\mathcal{R}$ is uniquely determined. Moreover, the new state, in which $M$ places the system, depends not only on the arrangement of $M$, but also on the result of $M$ (which could not be predicted causally in the original state) ---because the value of $\mathcal{R}$ in the new state must actually be equal to this $M$-result''. Or in Dirac's words: ``When we measure a real dynamical variable $\xi$, the disturbance involved in the act of measurement causes a jump in the state of the dynamical system. From physical continuity, if we make a second measurement of the same dynamical variable $\xi$ immediately after the first, the result of the second measurement must be the same as that of the first'' \cite[p. 36]{Dirac74}.

The introduction of the collapse within the axiomatic formulation of the theory has produced a series of paradoxes of which entanglement has not been the exception. The EPR {\it Gedeankenexperiment} exposes this paradox in what seems to be a non-local influence when measuring one of the particles on the other distant entangled partner. Indeed, if one accepts the orthodox interpretation of QM according to which the measurement of a quantum superposition induces a ``collapse'' to only one of its terms, Einstein Podolsky and Rosen then show that there then seems to exist a super-luminous transfer of information from one particle to the other distant partner. Einstein was of course clearly mortified by this seemingly non-local ``quantum effect'' which he called {\it ``spukhafte Fernwirkung''}, translated later as ``spooky action at a distance''. Once the entangled particles are separated, all their properties still remain {\it indeterminate}. But, the moment we perform a measurement of an observable in one of the particles we also find out instantaneously what is the value of the distant partner ---in case we would choose to measure the same observable. Thus, the ``collapse'' of one of the particles also produces the ``collapse'' of the other distant entangled particle. Every time we measure an observable in one of the particles, the other particle ---as predicted by QM--- will be found to possess a strictly correlated value.\footnote{Let us remark that observability is used in this case a sufficient condition to define reality itself. There is involved here a two sided definition of what accounts for physical reality, either in terms of computing the certainty of an outcome (= real) or by observing an outcome which was uncertain but became actual (= real).} Einsetin was not only against this ``spooky action'', he was also against the addition of a subjectively produced ``collapse''. In this respect, Einstein is quoted by Everett \cite[p. 7]{OsnaghiFreitasFreire09} to have said that he ``could not believe that  a mouse could bring about drastic changes in the universe simply by looking at it''. Schr\"odinger would also criticize the addition of the collapse to the theory:

\begin{quotation}
\noindent {\small``But jokes apart, I shall not waste the time by tritely ridiculing the attitude that the state-vector (or wave function) undergoes an abrupt change, when `I' choose to inspect a registering tape. (Another person does not inspect it, hence for him no change occurs.) The orthodox school wards off such insulting smiles by calling us to order: would we at last take notice of the fact that according to them the wave function does not indicate the state of the physical object but its relation to the subject; this relation depends on the knowledge the subject has acquired, which may differ for different subjects, and so must the wave function.'' \cite[p. 9]{OsnaghiFreitasFreire09}} \end{quotation}


\subsection{EPR's Conclusion: Incompleteness and Hidden Variables}

As it is well known, the conclusion of the EPR paper is that QM is an incomplete theory; it does not consider all the {\it elements of physical reality} the theory should talk about. In a nut shell, the line of reasoning in order to end up with such a fatal veredict for quantum theory runs as follows: 

\begin{enumerate}
\item[I.] A {\it complete theory} is one which takes into account all its {\it elements of physical reality}. 

\item[II.]  QM presents a limit, due to to Heisenberg's uncertainty relations, to the knowledge of complementary properties (of the same quantum system).

\item[III.] If one accepts as a {\it sufficient condition} the proposed definition of an {\it element of physical reality}, then one can argue that incompatible properties of the same quantum system must be regarded as being all {\it elements of physical reality}, simultaneously. 

\item[IV.] If incompatible complementary properties are elements of physical reality, this means they possess a definite value previous to measurement, thus contradicting Heisenberg's relations ---a cornerstone of orthodox QM itself.

\item[V.] Since QM is incapable of considering all the {\it elements of physical reality} (i.e., the definite valued properties) of the separated systems under study simultaneously (due to Heisenberg's relations), the theory is incomplete. 
\end{enumerate}

\noindent Einstein, Podolsky and Rosen conclude that QM is incomplete. However, they also maintain that it should be possible to ``complete the theory''. A new theory that considers ---unlike quantum theory--- all elements of physical reality present within an EPR type experiment could be developed in the future. It is with this hopeful wish that they end the paper \cite[p. 555]{Schr35b}: ``While we have thus shown that the wave function does not provide a complete description of the physical reality, we left open the question of whether or not such a description exists. We believe, however, that such a theory is possible.'' Thus, according to EPR, a description in terms of separable systems with definite valued properties should be, in principle, possible to develop. This has been understood in the literature ---due to Bell's reading \cite{Bell64}--- as implying the existence of a ``hidden variable theory'' which is able to account for an {\it actual state of affairs} and restore in this way an actualist description of nature. Unfortunately, due to Bohr's influence within 20th Century physics, all this deeply interesting debate regarding the definition of physical reality and correlations in QM was silenced for almost half a century.

\section{Particle Metaphysics Within the Contemporary Definition of\\ Quantum Entanglement}

As remarked by Jeffrey Bub \cite{Bub17}, ``[...] it was not until the 1980s that physicists, computer scientists, and cryptographers began to regard the non-local correlations of entangled quantum states as a new kind of non-classical resource that could be exploited, rather than an embarrassment to be explained away.''  The reason behind this shift in attitude towards {\it entanglement} is an interesting one. As Bub continues to explain: ``Most physicists attributed the puzzling features of entangled quantum states to Einstein's inappropriate `detached observer' view of physical theory, and regarded Bohr's reply to the EPR argument (Bohr, 1935) as vindicating the Copenhagen interpretation. This was unfortunate, because the study of entanglement was ignored for thirty years until John Bell's reconsideration of the EPR argument (Bell, 1964).'' Indeed, after the triumph of Bohr in the ``EPR battle'' \cite{Bohr35, EPR}, the notion of entanglement was almost completely erased by the orthodox community of physicists under the Copenhagen spell. This was until an Irish researcher called John Stewart Bell working at the Conseil Europ\'een pour la Recherche Nucl\'eaire (CERN), wrote in 1964 a paper entitled {\it On the Einstein-Podolsky-Rosen Paradox}. In this paper he was able to derive a set of statistical inequalities that restricted the correlations described by any classical local-realistic theory \cite{Bell64}. But the true breaking point for the recognition of quantum entanglement and the possibilities it implied for quantum information processing was the unwanted  result of the famous experiment performed in Orsay at the very beginning of the 1980s by Alain Aspect, Philippe Grangier and Gerard Roger \cite{AGR81}. The result was that the Bell inequality was violated by pairs of entangled spin ``particles''. As a consequence, against Einstein and Bell's physical intuition, the possibility for classical theories to account for such experience was completely ruled out. 

The experiment designed by Aspect and his team ---repeated countless times up to the present \cite{Bernien13, Hensen15}--- could not be described by any classical local-(binary) realistic\footnote{Even though the original term is ``realistic'', we prefer to add ``binary'' for reasons that will become evident in the forgoing part of the paper.} theory. The experiment was also a sign that entanglement had to be taken seriously. It was only then that quantum computation, quantum cryptography and quantum teleportation, were developed by taking {\it entanglement} as a resource \cite{Bub17}. The new notion began to rapidly populate the journals, labs and research institutions all around the world. The technological era of quantum information processing had woken up from its almost half century hibernation. An hibernation, let us not forget, mainly due to the uncritical attitude of the majority of physicists who believed that Bohr had already solved everything ---and there was no reason to engage in metaphysical questions regarding physical reality. 

With the advent of the new millennia the era of quantum information processing became rapidly one of the main centers of research and technology around the globe. It then became necessary to reach a consensus regarding the definition of quantum entanglement ---a notion which stood at the basis of all possibilities of technical analysis and development. Its definition had been clearly established by Schr\"odinger on the basis of two main notions: {\it separability} and {\it pure states}. But while Schr\"odinger was quite uneasy with ``collapses'' and the reference of the theory, the contemporary research community has simply embraced the definition together with its problematic presuppositions. A good example is the explanation of entanglement provided by Mintert et al.:  
\begin{quotation}
\noindent {\small ``Composite quantum systems are systems that naturally decompose into two
or more subsystems, where each subsystem itself is a proper quantum system. Referring to a decomposition as `natural' implies that it is given in an obvious fashion due to the physical situation. Most frequently, the individual susbsystems are characterized by their mutual distance that is larger than the size of a subsystem. A typical example is a string of ions, where each ion is a subsystem, and the entire string is the composite system. Formally, the Hilbert space ${\mathcal H}$ associated with a composite, or multipartite system, is given by the tensor product ${\mathcal H_1} \otimes ... \otimes {\mathcal H_N}$ of the spaces corresponding to each of the subsystems.'' \cite[p. 61]{Mintert09}}
\end{quotation}

\noindent Already this seemingly ``natural'' introduction to QM makes implicit use of an interpretation of the orthodox quantum formalism which is not ``obvious'' nor ``self evident'' at all. First, it implies the idea that Hilbert spaces can adequately represent `physical systems'; small elementary particles such as ions. And secondly, it also implies that such particles inhabit space-time, that one can make reference to distances and that the subspaces ---which can generate the whole Hilbert space--- describe `subsystems' ---i.e., a part of the original system. The problem is that none of this atomist ideas can be easily related to the quantum formalism. Assuming the metaphysics of particles as a ``common sense'' {\it given} of physical representation, the story of entanglement is then told in the following manner.  

In general, the Hilbert space associated with a composite system is given by the tensor product $\mathcal{H}_1\otimes\ldots\otimes \mathcal{H}_n$ of the spaces corresponding to each of the subsystems. The idea is that we should focus on a finite dimensional bipartite quantum system described by the Hilbert space $\mathcal{H}=\mathcal{H}_1\otimes \mathcal{H}_2$. After introducing {\it separability}, another essential element enters the scene, namely, the notion of {\it pure state}. The orthodox account of pure states rests in the following operational definition: If a quantum system is prepared in such way that one can devise a maximal test yielding with certainty a particular outcome, then it is said that the quantum system is in a \emph{pure state}. It is then stated that the pure state of a quantum system is described by a unit vector in a Hilbert space which in Dirac's notation is denoted by $|\psi \rangle$.\footnote{As discussed in \cite{daCostadeRonde16, deRondeMassri18b} this definition is ambiguous due to the non-explicit reference to the basis in which the vector is written. It is this ambiguity which, in turn, mixes the notion of `state of a system' and `property of a system'.} Assume now that each subsystem is prepared in the following pure states $|\psi\rangle$ and $|\psi'\rangle$. The state of the composite system is then $|\psi\rangle\otimes|\psi'\rangle$. Suppose that one had access to only one of the subsystems at a time. Then, after a measurement of any local observable $A\otimes\mathbb{I}$ on the first subsystem, (where $A$ is a hermitian operator acting on $\mathcal{H}_1$, and  $\mathbb{I}$ is the identity acting on $\mathcal{H}_2$), the state of the first subsystem will be projected onto an eigenstate of $A$, but the state of the second subsystem will remain unchanged. If later on, one performs a second local measurement, now on the second subsystem, it will yield a result that is completely independent of the result of the first measurement pertaining to the first subsystem. Hence, the measurement outcomes on the two subsystems are uncorrelated between each other and only depend on their own subsystem states.

In general, depending on the basis, a pure state in $\mathcal{H}$ is given by a superposition of pure states, $|\varphi\rangle=\sum a_i|\psi\rangle_i\otimes |\psi'\rangle_i$.
For a local operator on the first subsystem, the expected value is
\[
\mbox{Tr}(A\otimes\mathbb{I}|\varphi\rangle\langle \varphi|)=
\mbox{Tr}_1(A\rho_1),\quad \rho_1:=\mbox{Tr}_2(|\varphi\rangle\langle \varphi|),
\]
where $\mbox{Tr}_1$ and $\mbox{Tr}_2$ are the partial traces over the first and second subsystem and $\rho_1$ is the reduced density matrix of the first subsystem. Then, one can conclude that the state of the first subsystem is given by $\rho_1$ and the state of the second subsystem by $\rho_2$ (where $\rho_2:=\mbox{Tr}_1(|\varphi\rangle\langle \varphi|)$). However, the state of the composite system is different from $\rho_1\otimes\rho_2$. Moreover, if one performs a local measurement on one subsystem, this leads to a state reduction of the entire system state, not only of the subsystem on
which the measurement had been performed. Therefore, the probabilities for an outcome of a measurement on one subsystem are influenced by the measurements on the other distant subsystem. 
Thus, measurement results on subsystems are (classically) correlated.

\begin{definition}
States that can be written as a product of pure states are called \emph{product} or \emph{separable states}. The states which are not separable are then defined as \emph{entangled states}.
\end{definition} 

Now, let us consider mixed states. First of all, mixed product states $\rho\otimes\rho'$ do not exhibit correlations. However, a convex sum of different mixed product states, $\varrho=\sum p_i\rho_i\otimes\rho_i'$, where $p_i\ge 0$ and $\sum p_i=1$, yield correlated measurement results. In other words, there are local observables such that
\[
\mbox{Tr}(\varrho\,A\otimes B)\ne
\mbox{Tr}(\varrho\,A\otimes \mathbb{I})
\mbox{Tr}(\varrho\,\mathbb{I}\otimes B).
\]
These correlations can be described in terms of the classical probabilities $p_i$, and are therefore
considered classical. States of the form $\varrho$ are called \emph{separable mixed states}.  The other mixed states are called \emph{entangled mixed states}. Entangled states imply quantum correlations of measurements on different subsystems which cannot be described in terms of classical probabilities. 

The notions of {\it purity} and {\it separability} play an essential role in the orthodox definition of quantum entanglement. A first problem we find is that entanglement is defined in contradistinction to the notion of separability. According to orthodoxy, that which is not separable is entangled. This purely negative definition does not provide any direct intuitive grasp to the meaning of entanglement. Another problem, as remarked by Li and Quiao \cite[p. 1]{LiQuiao18}, is that the definition of separable is not even clear when considering density operators: ``A fundamental problem in the study of entanglement is the determination of the separability of quantum states. [...] The entanglement (non-separability) criterion for pure state is clear, by virtue of Schmidt or high order singular value decomposition for any-number-partite system. However, none of the existing criteria for the separability of finite dimensional mixed states are satisfactory by far. They are generally either sufficient and necessary, but not practically usable; or easy to use, but only necessary (or only sufficient).'' We believe that these difficulties are just an expression of a much deeper problem: the improper definition of quantum entanglement itself which has been grounded on strong metaphysical presuppositions which are not only alien to the mathematical formalism of the theory, but even worse, explicitly contradict it. In the following, taking as a standpoint the orthodox formalism of QM alone, we will argue against the distinctions introduced by the unjustified application of the metaphysics of space-time entities.

\section{Beyond Particle Metaphysics in QM} 

The metaphysical picture provided by atomist Newtonian mechanics remains a very heavy burden for today's quantum physics. Newton's metaphysical representation of the world has become the ``common sense'' of our time, advocated by many like a dogma that cannot be questioned. But the worst part of this situation comes from scientists ---and even philosophers--- who do not even seem to acknowledge that the Newtonian atomist picture {\it is} in fact a specific metaphysical picture with a long history going back to the atomism proposed by Democritus and Leucippus, and not an ``obvious'' or ``self evident'' unescapable way to talk about reality. Indeed, the notions of `state' and `system' applied to the quantum formalism imply the presupposition that QM talks about individual physical entities. This was the first intuition of the early atomic theory which found its origin in the Democritean theory of atoms. As pointed out by Heisenberg \cite[p. 218]{Castellani98} himself: ``The strongest influence on the physics and chemistry of the past [19th] century undoubtedly came from the atomism of Democritos. This view allows an intuitive description of chemical processes on a small scale. Atoms can be compared with the mass points of Newtonian mechanics, and from this a satisfactory statistical theory of heat was developed. [...] the electron, the proton, and possibly the neutron could, it seemed, be considered as the genuine atoms, the indivisible building blocks, of matter.'' In this respect, the present approach towards quantum theory remains not only dogmatic but also unscientific in a fundamental point. Going explicitly against what the formalism explicitly says, orthodoxy has implicitly assumed that we already know what the theory is talking about. This is exactly the opposite standpoint of scientific research which begins by humbly accepting the unknown. Scientific research is not a process of justification of our prejudices regarding our ``common sense'' picture of reality, it quite the opposite. The acceptance of the unknown and the effort to expand our understanding of reality. In the following we will show it is in fact possible to understand the orthodox formalism of QM without making any reference to ``tiny particles'' and strange ``collapses''.

\subsection{Beyond Purity, Collapses and Binary Existence}  

EPR's definition of {\it element of physical reality}, which has played a central role also within the definition of entanglement, is intrinsically related to the definition of pure state. It is only pure states which allow us to consider observables as {\it elements of physical reality} (in the EPR sense). Pure states guarantee the existence of an observable which is {\it certain} (probability equal to 1) if measured. It is only pure states which allow an interpretation of a quantum observable in terms of an {\it actual property}; i.e., a property that will yield the answer {\it yes} when being measured. At the opposite side, mixed states do not describe observables which, when measured, will be certain. When considering mixed states, all observables become {\it uncertain}; they all possess a probability which pertains to the open interval $(0,1)$. These properties are referred to in the literature as {\it indeterminate}  or {\it potential} properties. Indeterminate properties might, or might not become actualized in a future instant of time; they are {\it uncertain} properties which cannot be considered as elements of physical reality (in the EPR sense). However, as we mentioned above, a pure state in $\mathcal{H}$ is, in general, given by a superposition of pure states, $|\varphi\rangle=\sum a_i|\psi\rangle_i\otimes |\psi'\rangle_i$. It is at this point that the empiricist-positivist understanding of physics ---mainly as a formal scheme capable of describing observations--- has deeply influenced the need to introduce a `projection postulate' that would allow to transform quantum superpositions into single outcomes ---which is, as argued by orthodoxy, what we really observe. This postulate, added to the axiomatic formulation of the theory in a completely {\it ad hoc} manner, is then interpreted as a real ``collapse''; i.e., a physical process. The collapse takes place each time an observer decides to perform a measurement ---which, in turn, has also lead to the creation of the infamous measurement problem. All these (metaphysical) additions and restrictions grounded on certainty, purity and collapses are completely foreign to the orthodox Hilbert mathematical structure. The difficulties to talk about `systems' composed by `properties' are most extremely exposed by the superposition principle and quantum contextuality (see \cite{deRonde17a, deRondeMassri16}). But also, as remarked by Dennis Dieks \cite[p. 120]{Dieks10}: ``Collapses constitute a process of evolution that conflicts with the evolution governed by the Schr\"{o}dinger equation. And this raises the question of exactly when during the measurement process such a collapse could take place or, in other words, of when the Schr\"{o}dinger equation is suspended. This question has become very urgent in the last couple of decades, during which sophisticated experiments have clearly demonstrated that in interaction processes on the sub-microscopic, microscopic and mesoscopic scales collapses are never encountered.'' In the last decades, the experimental research seems to confirm there is nothing like a ``real collapse'' suddenly happening when measurement takes place. Unfortunately, as Dieks \cite{Dieks18} also acknowledges: ``The evidence against collapses has not yet affected the textbook tradition, which has not questioned the status of collapses as a mechanism of evolution alongside unitary Schr\"odinger dynamics.'' 

Leaving aside metaphysical pictures and focusing only on the orthodox formalism of QM itself it is possible to derive an important set of consequences which can guide our search for a new comprehension of what the theory of quanta is really talking about. To take the formalism seriously means for us to seek for an objective set of concepts which are grounded on the mathematical structure of the formalism itself. In particular, as we have argued elsewhere \cite{deRondeMassri16}, the key to understand the objective aspect of the mathematical formalism of QM is not hidden, it is exposed in the invariant structure of the theory. As Born himself reflected: \cite{Born53}: ``the idea of invariant is the clue to a rational concept of reality, not only in physics but in every aspect of the world.'' In physics, invariants are quantities having the same value for any reference frame. The transformations that allow us to consider the physical magnitudes from different frames of reference have the property of forming a group. It is this feature which allows us to determine what can be considered {\it the same} according to a mathematical formalism. In the case of classical mechanics invariance is provided via the Galilei transformations, while in relativity theory we have the Lorentz transformations. In QM the invariance of the theory is exposed by no other than Born's famous rule.

\smallskip

\smallskip

\noindent {\it
{\bf Born Rule:} Given a vector $\Psi$ in a Hilbert space, the following rule allows us to predict the average value of (any) observable $P$. 
$$\langle \Psi| P | \Psi \rangle = \langle P \rangle$$
This prediction is independent of the choice of any particular basis.}

\smallskip

\smallskip

\noindent This rule, which provides the invariant structure of the theory, points implicitly to the way in which physical reality should be conceived according to QM. Taking distance from the famous Bohrian prohibition to consider physical reality beyond the theories of Newton and Maxwell, we have proposed the following extended definition of what can be naturally considered ---by simply taking into account the mathematical invariance of the Hilbert formalism--- as a generalized element of (quantum) physical reality (see \cite{deRonde16a}).

\smallskip
\smallskip

\noindent {\it {\bf Generalized Element of Physical Reality:} If we can predict in any way (i.e., both probabilistically or with certainty) the value of a physical quantity, then there exists an element of reality corresponding to that quantity.}

\smallskip
\smallskip

\noindent This redefinition implies a deep reconfiguration of the meaning of the quantum formalism and the type of predictions it provides. It also allows to understand Born's probabilistic rule in a new light; not as providing information about a (subjective) measurement result, but instead, as providing objective information of a theoretically described (potential) state of affairs. Objective probability does not mean that particles behave in an intrinsically random manner. Objective probability means that probability characterizes a feature of the conceptual representation accurately and independently of any subjective choice or particular observation. This account of probability allows us to restore a representation in which the state of affairs is detached from the observer's choices to measure (or not) a particular property ---just like Einstein requested. This means that within our account of QM ---contrary to the orthodox viewpoint---, the Born rule always provides complete knowledge of the state of affairs described quantum mechanically; in cases where the probability is equal to 1 and also in cases in which probability is different to 1. In other words, both {\it pure states} and {\it mixed states} provide {\it maximal knowledge} of the (quantum) state of affairs. Since there is no essential mathematical distinction, both type of states have to be equally considered; none of them is ``less real'', or ``less well defined'' than the other according to the mathematics. Thus, it is not necessary at all to distinguish between pure states and mixed states. Or, in other words, there is nothing within the mathematical formalism which allows to build such a distinction.\footnote{This point has been already addressed by David Mermin in \cite[Sect. VII]{Mermin98}.} 

An explicit account of the orthodox quantum formalism without reference to collapses or particles has been already put forward in the categorical terms the logos approach to QM. Following \cite{deRondeMassri18a}, let $\mathcal{C}$ be a category and let $C$ be an object in $\mathcal{C}$. Let us define the category over $C$ denoted $\mathcal{C}|_C$. We assume that the reader is familiar with the definition of a \emph{category}. Objects in $\mathcal{C}|_C$ are given by arrows to $C$, $p:X\rightarrow C$,  $q:Y\rightarrow C$, etc. Arrows $f:p\rightarrow q$
are commutative triangles,
\[
\xymatrix{
X\ar[rr]^f\ar[dr]_p& &Y\ar[dl]^q\\
&C
}
\]

\noindent For example, let $\mathcal{S}ets|_\mathbf{2}$ be the category of sets
over $\mathbf{2}$, where $\textbf{2}=\{0,1\}$ and $\mathcal{S}ets$ is
the category of sets.
Objects in $\mathcal{S}ets|_\mathbf{2}$
are functions from a set to $\{0,1\}$
and morphisms are commuting triangles, 
\[
\xymatrix{
X\ar[rr]^f\ar[dr]_p& &Y\ar[dl]^q\\
&\{0,1\}
}
\]
In the previous triangle, $p$ and $q$ are objects of 
$\mathcal{S}ets|_\mathbf{2}$
and $f$ is a function satisfying $qf=p$.

Our main interest is the category $\mathcal{G}ph|_{[0,1]}$ of graphs over the interval $[0,1]$. The category $\mathcal{G}ph|_{[0,1]}$ has very nice categorical properties \cite{quasitopoi, graphtheory}, and it is a \emph{logos}.  Let us begin by reviewing some properties of the category of graphs. A \emph{graph} is a set with a reflexive symmetric relation. The category of graphs extends naturally the category of sets and the category of aggregates (objects with an equivalence relation). A set is a graph without edges. An {\it aggregate} is a graph  in which the relation is transitive. More generally, we can assign to a category a graph, where the objects are the nodes of the graph and there is an edge between $A$ and $B$ if $\hom(A,B)\neq\emptyset$.
Given that in a category we have a composition law, the resulting graph is an aggregate.

\begin{definition}
We say that a graph $\mathcal{G}$ is \emph{complete} if there is an edge between two arbitrary nodes. A \emph{context} is a complete subgraph (or aggregate) inside $\mathcal{G}$. A \emph{maximal context} is a context not contained properly in another context. If we do 
not indicate the opposite, when we refer to contexts we will be implying maximal contexts.
\end{definition}

\noindent For example, let $P_1,P_2$ be two elements of a graph $\mathcal{G}$. 
Then, $\{P_1, P_2\}$ is a contexts if $P_1$ is related to $P_2$, $P_1\sim P_2$. Saying differently, if there exists an edge between $P_1$ and $P_2$. In general, a collection of elements $\{P_i\}_{i\in I}\subseteq \mathcal{G}$ determine a {\it context} if $P_i\sim P_j$ for all $i,j\in I$. Equivalently, if the subgraph with nodes $\{P_i\}_{i\in I}$ is complete. 

Given a Hilbert space $\mathcal{H}$, 
we can define naturally a graph $\mathcal{G}=\mathcal{G}(\mathcal{H})$
as follows. Following [{\it Op. cit.}] the nodes are interpreted as {\it immanent powers} and there exists an edge between  $P$ and $Q$ if and only if $[P,Q]=0$. This relation makes $\mathcal{G}$ a graph (the relation is not transitive). We call this relation {\it quantum commuting relation}.

\begin{theo}
Let $\mathcal{H}$ be a Hilbert space and let $\mathcal{G}$
be the graph of immanent powers with the commuting relation given by QM. 
It then follows that: 
\begin{enumerate}
\item The graph $\mathcal{G}$ contains all the contexts. 
\item Each context is capable of generating the whole graph $\mathcal{G}$.
\end{enumerate}
\end{theo}
\begin{proof}
See \cite{deRondeMassri18b}.
\qed
\end{proof}

\smallskip
\smallskip

As we mentioned earlier, an object in $\mathcal{G}ph|_{[0,1]}$ consists in  a map $\Psi:\mathcal{G}\rightarrow [0,1]$, where $\mathcal{G}$ is a graph. Then, in order to provide a map to the graph of immanent powers, we use the Born rule. To each \emph{power} $P\in\mathcal{G}$, we assign through the Born rule  the number $p=\Psi(P)$, where $p$ is a number between $0$ and $1$ called \emph{potentia}. As discussed in detail in \cite{deRondeMassri18a}, we call this  map $\Psi:\mathcal{G}\rightarrow [0,1]$ a {\it Potential State of Affairs} (PSA for short). Summarizing, we have the following:

\begin{definition}
Let $\mathcal{H}$ be Hilbert space and let $\rho$ be a density matrix.
Take $\mathcal{G}$ as the graph of immanent powers with the quantum commuting relation. 
To each immanent power $P\in\mathcal{G}$ apply the Born rule to get the number $\Psi(P)\in[0,1]$, which is called the potentia (or intensity) of the power $P$.  Then, $\Psi:\mathcal{G}\rightarrow [0,1]$
defines an object in $\mathcal{G}ph|_{[0,1]}$. We call this map a \emph{Potential State of Affairs}.
\end{definition}

Intuitively, we can picture a PSA as a table,
\[
\Psi:\mathcal{G}(\mathcal{H})\rightarrow[0,1],\quad
\Psi:
\left\{
\begin{array}{rcl}
P_1 &\rightarrow &p_1\\
P_2 &\rightarrow &p_2\\
P_3 &\rightarrow &p_3\\
  &\vdots&
\end{array}
\right.
\]

The introduction of {\it intensive valuations} allows us to derive a non-contextuality theorem that is able to escape Kochen-Specker contextuality \cite{deRondeMassri18a}. 

\begin{theo}
The knowledge of a PSA $\Psi$ is equivalent to the knowledge of the density matrix $\rho_{\Psi}$. In particular, if $\Psi$
is defined by a normalized vector $v_{\Psi}$, $\|v_{\Psi}\|=1$, then we can recover the vector from $\Psi$.

\end{theo}
\begin{proof}
See \cite{deRondeMassri18b}.
\qed
\end{proof}
\smallskip
\smallskip

Notice that our mathematical representation is objective in the sense that it relates, in a coherent manner and without internal contradictions, the multiple contexts (or aggregates) to the whole PSA. Contrary to the contextual (relativist) Bohrian ``complementarity solution'' grounded on his doctrine of classical concepts, there is in this case no need of a (subjective) choice of a particular context in order to define the ``physically real'' state of affairs. The state of affairs is described completely by the whole graph (or $\Psi$), and the contexts bear an invariant (objective) existence independently of any (subjective) choice. Let us remark that `objective' is not understood as a synonym of `real', but rather as providing the conditions of a theoretical representation in which all subjects are {\it detached} from the course of events. Contrary to Bohr's claim, in our account of QM, individual subjects are not considered as actors. Subjects are humble spectators and their choices do not change the objective representation provided by the theory.

The introduction of a ``collapse'' scrambles the objective quantum theoretical representation with subjective epistemic observations. A direct consequence of this scrambling is that the change in our (subjective) knowledge also changes the theoretical description of (objective) reality itself.\footnote{This same scrambling takes place in the case of quantum contextuality and the so called ``basis problem''. See for a detailed analysis: \cite{deRonde16c}.} However, if we simply accept that quantum probability is not making reference to measurement outcomes, but instead characterizes an objective (intensIve) feature of the state of affairs represented by QM, then there is no need of considering epistemic observations within the theoretical representation. In such case, as we have shown above, we can restore the objective nature of the phenomena without the intromission of epistemic knowledge within the description of objective quantum reality. The price to pay is to accept that QM talks about reality in a very different way than classical physics.

\subsection{Beyond Space-Time Separability}

In \cite{Aerts84a}, Diederik Aerts discussed the EPR paradox as an {\it ad absurdum} proof making explicit ``the missing elements of reality in the description by quantum mechanics of separated physical systems.'' His deep and interesting analysis of the EPR paper can be nicely summarized in the following passage: 
\begin{quotation}
\noindent {\small ``[...] what E.P.R. show is the following

\smallskip

{\it Quantum mechanics describes correctly and in a complete way separated systems.

$\Rightarrow$ Quantities that are not compatible can have simultaneous reality.

$\Rightarrow$ Quantum mechanics is not complete.}

\smallskip

\noindent From this they can conclude that

\smallskip

{\bf D} {\it the quantum mechanical description of separated systems is not correct or not complete,} 

\noindent or

{\bf E} {\it quantities that are not compatible can have simultaneous reality and hence 
quantum mechanics is not complete.}

\smallskip

\noindent E.P.R. mention in the beginning of their paper that they suppose quantum 
mechanics to be correct. Hence they then also suppose quantum mechanics to 
give a correct description of separated systems. Two alternatives remain in this 
case: the quantum mechanical description of separated systems is incomplete or 
quantum mechanics is incomplete in the sense that quantities that are not 
compatible can have simultaneous reality. But in any case quantum mechanics is 
not complete. Hence E.P.R. can conclude that if quantum mechanics is correct, 
then it is not complete. What we showed in this paper is that the incompleteness 
arrives from {\bf D} and not from {\bf E}. Quantum mechanics is incomplete because it does 
not give a complete description of separated systems.''\cite[p. 427]{Aerts84a}}
\end{quotation}


\noindent In \cite{Aerts84b}, Aerts proposed a solution to the paradox. Holding fast to the operational definition of an {\it element of physical reality}, his solution attempted to change QM ``in order to describe adequately separated systems''. Like Einstein, Aerts seemed willing to retain {\it separability} even at the cost of transforming a mathematical formalism which captured ---in a quantitative manner--- an immense number of phenomena. However, taking once again the orthodox formalism as a standpoint, there is another quite obvious solution which we have investigated within the logos approach, namely, to simply abandon {\it space-time separability} as a presupposition of the theory. We can certainly drop the principle since, as already remarked by Einstein himself \cite[p. 172]{Born71}, ``quantum theory nowhere makes explicit use of this requirement.'' In fact, just to add to the list, there is another obvious aspect of the formalism which precludes the possibility to understand Hilbert spaces as making references to `systems'  and `subsystems' and makes explicit the formal-conceptual incoherency put forward by the orthodox ``minimal'' interpretation of QM. While the equation of motion in classical mechanics can be expressed in ${\cal R}^3$ allowing an interpretation in tridimensional Euclidean space, QM works in a configuration space. The difference is essential when attempting to consider existents within space. Classically, if we add two systems, the properties are summed. Given two systems with a number of properties, $R$ and $R'$, respectively; their joint consideration is just the sum of the properties of each system, namely, $R+R$. A paradigmatic example is the completely inelastic crash of two systems. While before the crash the two particles are separated and their mass are $m$ and $m'$, and their velocities are $v$ and $v'$, respectively; after the crash they become a (non-separable) single system of mass $m+m'$ with a common velocity $v_f$. The essential property characterizing the two systems ---namely, their mass--- becomes nothing else than the sum of masses. But, as we know, there is an essential difference when considering the addition of `systems' (vector spaces) in QM. If we take two rays which intersect each other, in terms of classical set theory, the addition of the rays is just the two rays; however, in terms of vector spaces the addition of two rays (now considered as subspaces) is more than just their sum, it is the whole plane generated by the two rays (see also the analysis provided by Rob Griffiths in \cite[Sect. 2]{Griffiths02}). In QM, the new possibilities considered by the addition of systems are not just the sum of the previous subsystems, they are much more.\footnote{The logos approach provides an intuitive understanding of what is going on in terms of the capabilities of an apparatus: adding two apparatuses allows many more possibilities than just the reductionistic sum of their previous possibilities.}


\section{In Search of a New Definition of Quantum Entanglement} 

In the history of science, many times, we physicists, have been confronted with ``spooky situations''. The ``spookiness'' always comes from the lack of understanding. Incomprehension of the unknown is always frightening. Just to give an example between many, the phenomena of electricity and magnetism were regarded as ``magical'' since the origin of humanity itself. Pieces of stone attracting each other without any material contact can be indeed ``spooky'', not to talk about lightnings coming from the sky. For a long period of time the only explanation was that some God, like Zeus, was the one responsible for this process. This was until one day physicists were finally able to create a theory called electromagnetism which explained all these different phenomena in both a qualitative and quantitative manner. Physicists were even able to find out that these apparently different phenomena could be quantified in terms of a unified mathematical formalism. But it was only through the conceptual representation of `fields', that we could finally grasp a deep qualitative understanding of the phenomena. In the end, electricity and magnetism were two sides of the same represented physical reality. Suddenly, the spookiness had disappeared. 

Nobel laureate Steven Weinberg \cite{Weinberg}, when discussing about quantum entanglement, has argued that: ``What is surprising is that when you make a measurement of one particle you affect the state of the other particle, you change its state!'' This conclusion is indeed spooky, since we never observe in our macroscopic world that objects behave in such a strange manner. A table (or chair) in one place never affects the state of another distant table (or chair). If we do something to an object in a region of space $A$, there will be no instantaneous action produced on an object situated in a distant region $B$. But Weinberg's amazement is not detached from his biased metaphysical viewpoint, it is actually grounded on the ---unjustified--- assumption that QM talks about ``small particles'' which suddenly ``collapse'' when being observed. As we have argued extensively, one simply does not find within the formal structure of QM neither a {\it reference} to particles nor to any strange collapse process.\footnote{In this respect we might argue that the notion of particle, rather than helping us understand quantum phenomena seems to have played the role of an epistemological obstruction \cite{deRondeBontems11}.} 

Today, it is repeated that physics can only describe `systems' with definite `states' and `properties'. This understanding of physics restricted by the classical paradigm ---mainly due to Bohr's philosophy of physics supplemented by 20th Century positivism--- has blocked the possibility to advance in the development of a new conceptual scheme for QM. And exactly this is what David Deutsch \cite{Deutsch04} has rightly characterized as ``bad philosophy''; i.e., ``[a] philosophy that is not merely false, but actively prevents the growth of other knowledge.'' In the previous sections we have shown it is possible to advance in a representation of QM which avoids right from the start the atomist metaphysical picture inherited from Newtonian classical mechanics. It is, as we have explicitly shown, possible to define existence in a different way to that presupposed by atomist metaphysics. And this possibility allows to escape the spookiness involved when attempting to understand QM through the dogmatic viewpoint of ``small particles''. To summarize, our conclusion is that in case we accept the possibility of non-collapse interpretations of QM or even interpretations of the theory which do not make reference to `systems', the notion of quantum entanglement will require ---at least in these cases--- a new definition. One that goes beyond the metaphysical presuppositions and distinctions imposed by atomist metaphysics. The explicit derivation of this new definition exceeds the scope of the present paper which we leave for a future work.

\section{Conclusion} 

In this paper we have provided arguments against the orthodox definition of quantum entanglement. We have argued that the present definition discussed in the literature is grounded on purely metaphysical presuppositions which go, in fact, against what the formalism explicitly tells us. In this respect we have shown why the notion of {\it pure state} and that of {\it separability} are not essential to the mathematical structure of QM. The main conclusion of the paper is that, if we accept that QM does not talk about particles and that pure states are not essential, a new definition of quantum entanglement is required.

\section*{Acknowledgements} 

C. de Ronde would like to thank Don Howard for historical references. We would also like to thank Dirk Aerts, Massimiliano Sassoli de Bianchi, Raimundo Fernandez Moujan, Giuseppe Sergioli and Hector Freytes for related discussions. This work was partially supported by the following grants: FWO project G.0405.08 and FWO-research community W0.030.06. CONICET RES. 4541-12 and the Project PIO-CONICET-UNAJ (15520150100008CO) ``Quantum Superpositions in Quantum Information Processing''.

\end{document}